\documentclass[a4paper,11pt]{article}
\usepackage{pos}

\usepackage{array}
\usepackage{booktabs}
\usepackage{slashed}
\usepackage{subfig} 

\usepackage{tikz}
\usetikzlibrary{shapes.geometric, arrows,chains}
\usetikzlibrary{decorations.markings}

\usepackage{tikz-feynman}
\tikzfeynmanset{compat=1.1.0}
\tikzfeynmanset{warn luatex=false}

\title{Electroweak corrections to Higgs boson pair production: The quark channel}

\author[a,b,c]{Marco Bonetti}
\author[b]{Gudrun Heinrich}
\author*[b]{Philipp Rendler}
\author[d]{William J.\ Torres~Bobadilla}

\affiliation[a]{Institute for Theoretical Physics, University of Tübingen, Auf der Morgenstelle 14, 72076 Tübingen, Germany} 
\affiliation[b]{Institute for Theoretical Physics, Karlsruhe Institute
  of Technology (KIT), Wolfgang-Gaede-Str.~1, 76131 Karlsruhe, Germany}
\affiliation[c]{Institute for Astroparticle Physics, Karlsruhe Institute of Technology (KIT), 76344 Eggenstein-Leopoldshafen, Germany}
\affiliation[d]{Department of Mathematical Sciences, University of Liverpool, Liverpool L69 3BX, U.K.}
 
\emailAdd{marco.bonetti@uni-tuebingen.de}
\emailAdd{gudrun.heinrich@kit.edu}
\emailAdd{philipp.rendler@kit.edu}
\emailAdd{torres@liverpool.ac.uk}

\abstract{We present the mixed QCD-electroweak corrections to Higgs boson pair production in the quark--antiquark channel. The virtual amplitudes are computed fully analytically using the method of differential equations. We determine the integration constants by matching our expressions to the large mass expansion limit of the canonical integrals. We implement the results in the \texttt{POWHEG-BOX} framework for phenomenological studies. The corrections are found to have a significant impact on the shapes of differential cross sections, reaching up to $+10\%$ for the invariant mass distribution of the Higgs boson pair near the production threshold. This channel has not been considered before in calculations of the next-to-leading order electroweak corrections to Higgs boson pair production.}

\FullConference{Loops and Legs in Quantum Field Theory (LL2026)\\
12-17, April, 2026\\
Bayreuth, Germany\\}

\begin{document}
\maketitle

\section{Introduction}
Constraining the trilinear Higgs coupling is one of the primary goals of the High Luminosity phase of the LHC. Current constraints from the combined ATLAS and CMS analysis at the $95\%$ confidence level, $-0.71 < \lambda/\lambda_{SM} < 6.1$ \cite{CMS:2026nuu}, still allow for significant deviations from the Standard Model prediction, leaving room for beyond-the-Standard-Model physics. 

Higgs boson pair production is the primary channel for measuring the trilinear Higgs coupling, as the value of $\lambda$ directly affects both the total cross section and the shape of differential distributions. Precise theoretical predictions are therefore essential.

QCD corrections to Higgs boson pair production in gluon fusion have been computed at next-to-leading order (NLO) with full top mass dependence \cite{Borowka:2016ehy, Borowka:2016ypz, Baglio:2018lrj, Davies:2019dfy, Baglio:2020ini}, including parton shower matching \cite{Heinrich:2017kxx, Jones:2017giv, Heinrich:2019bkc, Bagnaschi:2023rbx, Davies:2025qjr, Alioli:2025xcu}, and up to $\text{N}^3\text{LO} +\text{N}^3\text{LL}$ with the highest order contribution calculated in the heavy-top limit \cite{Grazzini:2018bsd, deFlorian:2016uhr, Grigo:2015dia, Chen:2019lzz, Chen:2019fhs, AH:2022elh}. These calculations reduce the residual QCD scale uncertainty to approximately $\mathcal{O}(3\%)$. The top-mass renormalisation scheme uncertainty \cite{Bagnaschi:2023rbx, Baglio:2020wgt}, which remains the dominant uncertainty at the level of $\mathcal{O}(20\%)$, is currently under investigation and has been addressed in Refs. \cite{Jaskiewicz:2024xkd, Davies:2025ghl}. 

With the QCD uncertainties substantially reduced, NLO electroweak (EW) corrections become increasingly relevant. These corrections have been shown to contribute at the level of $\mathcal{O}(5\%)$ to the total cross section and up to $\mathcal{O}(30\%)$ for differential distributions \cite{Bi:2023bnq}. In addition to the full EW corrections, several gauge-invariant subsets have been studied in Refs. \cite{Borowka:2018pxx, Muhlleitner:2022ijf, Davies:2022ram, Davies:2023npk, Bizon:2024juq, Heinrich:2024dnz, Davies:2025wke, Bonetti:2025vfd, Bhattacharya:2025egw, Davies:2026wbx}.

These proceedings review Ref. \cite{Bonetti:2026cih}, which presented the first study of EW effects in Higgs boson pair production initiated by a quark--antiquark pair. Although these contributions are negligible at the level of the total cross section, they were shown to lead to significant distortions of differential distributions close to the production threshold. 

\section{Amplitude Calculation}
We consider Higgs boson pair production with an incoming quark--antiquark pair, 
\begin{align}
    q \overline{q} \to H H\, ,
\end{align}
with massless quarks of the first two generations, $q \in \{u, d, s, c\}$. Furthermore, we assume a diagonal CKM matrix. Under these assumptions, the Higgs bosons can only be produced through their couplings to EW gauge bosons. Representative Feynman diagrams contributing at leading order (LO) and NLO, separated into real-emission and virtual contributions, are shown in Tab.~\ref{tab:diagrams}. 

\begin{table}[tb]
\caption{Example Diagrams for LO and NLO $q\overline{q} \to HH$.}
\label{tab:diagrams}
\centering 
\begin{tabular}{>{\centering\arraybackslash}p{3cm}>{\centering\arraybackslash}p{5cm}>{\centering\arraybackslash}p{3cm}}
    \toprule
    Leading Order & \multicolumn{2}{c}{\hspace{1.5cm}Next-to-Leading Order QCD} \\
    & real emissions & virtual corrections \\
    \midrule
    \begin{tikzpicture}[baseline={([yshift=-0.5ex]hi.base)}, scale=0.8, transform shape]
    \begin{feynman}[inline = (hi)]
        \vertex (hi);
        \vertex [right = 0.6cm of hi] (ha);
        \vertex [right = 1.2cm of ha] (hb);
        \vertex [right = 0.6cm of hb] (hf);
        \vertex [above = 1.2cm of hi] (i1);
        \vertex [above = 0.6cm of ha] (a1);
        \vertex [above = 0.6cm of hb] (b1);
        \vertex [above = 1.2cm of hf] (f1);
        \vertex [below = 1.2cm of hi] (i2);
        \vertex [below = 0.6cm of ha] (a2);
        \vertex [below = 0.6cm of hb] (b2);
        \vertex [below = 1.2cm of hf] (f2);
        \diagram*{
            (i1) -- [fermion] (a1) -- [fermion] (a2) -- [fermion] (i2), 
            (a1) -- [boson] (b1) -- [boson] (b2) -- [boson] (a2), 
            (b1) -- [scalar] (f1), 
            (b2) -- [scalar] (f2),
        };
    \end{feynman}
\end{tikzpicture}  & 
    \begin{tikzpicture}[baseline={([yshift=-0.5ex]hi.base)}, scale=0.8, transform shape]
    \begin{feynman}[inline = (hi)]
        \vertex (hi);
        \vertex [right = 0.6cm of hi] (ha);
        \vertex [right = 1.2cm of ha] (hb);
        \vertex [right = 0.6cm of hb] (hf);
        \vertex [right = 0.3cm of hi] (hg1); 
        \vertex [right = 0.6cm of ha] (hg2); 
        \vertex [above = 1.2cm of hi] (i1);
        \vertex [above = 0.6cm of ha] (a1);
        \vertex [above = 0.6cm of hb] (b1);
        \vertex [above = 1.2cm of hf] (f1);
        \vertex [below = 1.2cm of hi] (i2);
        \vertex [below = 0.6cm of ha] (a2);
        \vertex [below = 0.6cm of hb] (b2);
        \vertex [below = 1.2cm of hf] (f2);

        \vertex [above = 0.9cm of hg1] (g1);
        \vertex [above = 1.5cm of hg2] (g2);
        \diagram*{
            (i1) -- [fermion] (a1) -- [fermion] (a2) -- [fermion] (i2), 
            (a1) -- [boson] (b1) -- [boson] (b2) -- [boson] (a2), 
            (b1) -- [scalar] (f1), 
            (b2) -- [scalar] (f2),
            (g1) -- [gluon] (g2),
        };
    \end{feynman}
\end{tikzpicture}  
    \begin{tikzpicture}[baseline={([yshift=-0.5ex]hi.base)}, scale=0.8, transform shape]
    \begin{feynman}[inline = (hi)]
        \vertex (hi);
        \vertex [right = 0.6cm of hi] (ha);
        \vertex [right = 1.2cm of ha] (hb);
        \vertex [right = 0.6cm of hb] (hf);
        \vertex [right = 0.3cm of hi] (hg1); 
        \vertex [right = 0.6cm of ha] (hg2); 
        \vertex [above = 1.2cm of hi] (i1);
        \vertex [above = 0.6cm of ha] (a1);
        \vertex [above = 0.6cm of hb] (b1);
        \vertex [above = 1.2cm of hf] (f1);
        \vertex [below = 1.2cm of hi] (i2);
        \vertex [below = 0.6cm of ha] (a2);
        \vertex [below = 0.6cm of hb] (b2);
        \vertex [below = 1.2cm of hf] (f2);

        \vertex [above = 1cm of hg1] (g1);
        \vertex [above = 1.5cm of hg2] (g2);

        \vertex [left = 0.5cm of i1] (ii1);
        \diagram*{
            (ii1) -- [gluon] (g1),
            (g2) -- [fermion] (g1) -- [fermion] (a1) -- [fermion] (a2) -- [fermion] (i2), 
            (a1) -- [boson] (b1) -- [boson] (b2) -- [boson] (a2), 
            (b1) -- [scalar] (f1), 
            (b2) -- [scalar] (f2)
        };
    \end{feynman}
\end{tikzpicture} 
    & 
    \begin{tikzpicture}[baseline={([yshift=-0.5ex]hi.base)}, scale=0.8, transform shape]
    \begin{feynman}[inline = (hi)]
        \vertex (hi);
        \vertex [right = 0.6cm of hi] (ha);
        \vertex [right = 1.2cm of ha] (hb);
        \vertex [right = 0.6cm of hb] (hf);
        \vertex [right = 0.3cm of hi] (hg); 
        \vertex [above = 1.2cm of hi] (i1);
        \vertex [above = 0.6cm of ha] (a1);
        \vertex [above = 0.6cm of hb] (b1);
        \vertex [above = 1.2cm of hf] (f1);
        \vertex [below = 1.2cm of hi] (i2);
        \vertex [below = 0.6cm of ha] (a2);
        \vertex [below = 0.6cm of hb] (b2);
        \vertex [below = 1.2cm of hf] (f2);

        \vertex [above = 0.9cm of hg] (g1);
        \diagram*{
            (i1) -- [fermion] (a1) -- [fermion] (a2) -- [fermion] (i2), 
            (a1) -- [boson] (b1) -- [boson] (b2) -- [boson] (a2), 
            (b1) -- [scalar] (f1), 
            (b2) -- [scalar] (f2),
            (g1) -- [gluon, quarter right] (ha),
        };
    \end{feynman}
\end{tikzpicture}  \\
    \bottomrule 
\end{tabular}
\end{table}

Since LO and real-emission amplitudes are generated automatically with \texttt{GoSam-3} \cite{Cullen:2011ac, GoSam:2014iqq, Braun:2025afl}, the remainder of this section focuses on the analytic calculation of the two-loop virtual corrections.

We generate all relevant Feynman diagrams using \texttt{QGRAF}~\cite{Nogueira:1991ex} and translate them into analytic expressions with \texttt{FORM~4.2}\footnote{Note, that \texttt{FORM~5.0} has since become available \cite{Davies:2026cci}.}~\cite{Ruijl:2017dtg} employing the Feynman rules of Ref.~\cite{Romao:2012pq}. The amplitudes are computed in both the unitary and Feynman gauge for the EW gauge bosons, finding agreement. These choices are particularly convenient, as they lead to the same set of master integrals.  

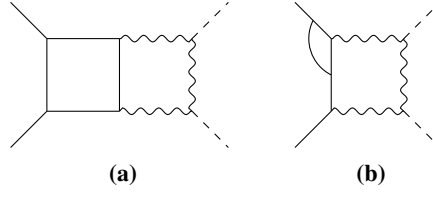
\begin{figure}[tb]
    \centering
    \subfloat[]{\begin{tikzpicture}[baseline={([yshift=-0.5ex]hi.base)}, scale=0.8, transform shape]
    \begin{feynman}[inline = (hi)]
        \vertex (hi);
        \vertex [right = 0.6cm of hi] (ha);
        \vertex [right = 1.2cm of ha] (hb);
        \vertex [right = 1.2cm of hb] (hc);
        \vertex [right = 0.6cm of hc] (hf);
        \vertex [right = 0.3cm of hi] (hg); 
        \vertex [above = 1.2cm of hi] (i1);
        \vertex [above = 0.6cm of ha] (a1);
        \vertex [above = 0.6cm of hb] (b1);
        \vertex [above = 0.6cm of hc] (c1);
        \vertex [above = 1.2cm of hf] (f1);
        \vertex [below = 1.2cm of hi] (i2);
        \vertex [below = 0.6cm of ha] (a2);
        \vertex [below = 0.6cm of hb] (b2);
        \vertex [below = 0.6cm of hc] (c2);
        \vertex [below = 1.2cm of hf] (f2);

        \vertex [above = 0.9cm of hg] (g1);
        \diagram*{
            (i1) -- (a1) -- (a2) -- (i2), 
            (a1) -- (b1) -- (b2) -- (a2), 
            (b1) -- [boson] (c1) -- [boson] (c2) -- [boson] (b2), 
            (c1) -- [scalar] (f1), 
            (c2) -- [scalar] (f2),
        };
    \end{feynman}
\end{tikzpicture} \label{fig:PL1}}
    \qquad
    \subfloat[]{\begin{tikzpicture}[baseline={([yshift=-0.5ex]hi.base)}, scale=0.8, transform shape]
    \begin{feynman}[inline = (hi)]
        \vertex (hi);
        \vertex [right = 0.6cm of hi] (ha);
        \vertex [right = 1.2cm of ha] (hb);
        \vertex [right = 0.6cm of hb] (hf);
        \vertex [right = 0.3cm of hi] (hg); 
        \vertex [above = 1.2cm of hi] (i1);
        \vertex [above = 0.6cm of ha] (a1);
        \vertex [above = 0.6cm of hb] (b1);
        \vertex [above = 1.2cm of hf] (f1);
        \vertex [below = 1.2cm of hi] (i2);
        \vertex [below = 0.6cm of ha] (a2);
        \vertex [below = 0.6cm of hb] (b2);
        \vertex [below = 1.2cm of hf] (f2);

        \vertex [above = 0.9cm of hg] (g1);
        \diagram*{
            (i1) -- (a1) -- (a2) -- (i2), 
            (a1) -- [boson] (b1) -- [boson] (b2) -- [boson] (a2), 
            (b1) -- [scalar] (f1), 
            (b2) -- [scalar] (f2),
            (g1) -- [quarter right] (ha),
        };
    \end{feynman}
\end{tikzpicture} \label{fig:PL2}}
    \caption{Top sectors of the virtual NLO. Straight lines indicate massless particles, curved lines EW bosons with mass $m_{W/Z}$ and dashed lines Higgs bosons with mass $m_H$.}
    \label{fig:top-sectors}
\end{figure}

To treat the $\gamma_5$ matrices appearing in quark--gauge-boson couplings, we employ naive dimensional regularisation (NDR) \cite{Chanowitz:1979zu}. In the present calculation, all $\gamma_5$ matrices occur on open fermion lines. Consequently, no traces containing $\gamma_5$ need to be evaluated in $D$ dimensions, thereby avoiding the ambiguities that may arise in NDR. This allows us to anticommute the $\gamma_5$ matrices along the fermion lines and attach them to the external spinors, yielding
\begin{align}
    \overline{v}(p_2)\, \mathcal{M}\,u(p_1) = \overline{v}_L(p_2)\, \mathcal{M}_L\, u_L(p_1) + \overline{v}_R(p_2)\, \mathcal{M}_R\, u_R(p_1)\, .
\end{align}
The remaining matrix elements $\mathcal{M}_{L/R}$ are free of $\gamma_5$ and can therefore be written in terms of a single form factor, 
\begin{align}
    \mathcal{M}_{L/R} \sim \mathcal{F}_{L/R} \slashed{p}_3\, .
\end{align}
For the evaluation of the Feynman integrals, we first use \texttt{Reduze~2} \cite{vonManteuffel:2012np} and \texttt{Kira~3} \cite{Maierhofer:2017gsa, Klappert:2019emp, Klappert:2020nbg, Klappert:2020aqs, Lange:2025fba} to apply integration-by-parts (IBP) relations and reduce them to a minimal set of $51$ master integrals associated with two top-level sectors shown in Fig.~\ref{fig:top-sectors}. The $45$ integrals related to the double-box topology, Fig.~\ref{fig:PL1}, were computed by some of us in Ref.~\cite{Bonetti:2025vfd} using canonical differential equations~\cite{Henn:2013pwa}, 
\begin{align}
    \mathrm{d}\mathbf{J} = \varepsilon \sum_{i = 1}^{51} \mathbb{A}_i \mathrm{dlog}(\alpha_i) \mathbf{J}\, , 
\end{align}
where $\mathbf{J}$ denotes the canonical integrals, $\alpha_i = \alpha_i(s, t, m_H^2, m_{W/Z}^2)$ are algebraic functions of the kinematic scales, $\mathbb{A}_i$ are rational matrices and $\varepsilon$ is the dimensional regulator. 

This form was achieved by first constructing a canonical basis using \texttt{DlogBasis} \cite{Henn:2020lye} for sectors up to six propagators, and a loop-by-loop approach \cite{Flieger:2022xyq} for the top sectors containing seven propagators. The corresponding differential equations were derived with the help of \texttt{LiteRed} \cite{Lee:2012cn} and \texttt{FiniteFlow} \cite{Peraro:2019svx} and subsequently brought into dlog form with the help of \texttt{Efforteless}~\cite{Antonela}. The boundary constants were fixed by evaluating the integrals in the large mass expansion, which corresponds to the limit $s, t, u, m_H^2 \ll m_{W/Z}^2$.

The remaining six integrals of the second topology, which is shown in Fig.~\ref{fig:PL2}, involve the same set of integration kernels and were brought into dlog form within the same framework. 

Having expressed all master integrals in terms of iterated integrals over logarithmic kernels, we perform additional basis rotations to a set of graded transcendental functions \cite{Chicherin:2021dyp, Gehrmann:2024tds}. These functions are constructed by determining linear relations among the integrals order by order in the dimensional regulator $\varepsilon$ using \texttt{FiniteFlow}. Applying these relations allows us to obtain a minimal set of independent functions at each order in $\varepsilon$ . 

Expressing the amplitude in terms of independent functions removes all artificial poles. The remaining UV poles are treated by renormalising the strong coupling constant $\alpha_s$ in the $\overline{MS}$ scheme, and the IR poles are removed using the Catani subtraction scheme \cite{Catani:1998bh}. 

\begin{figure}[tb]
    \centering 
        \begin{tikzpicture}[scale=0.35]
        \draw[->] (0,0) -- (20, 0) node[right] {$-t/m_H^2$};
        \draw[->] (0,0) -- (0, 20) node[above] {$s/m_H^2$};
        \draw[dotted] (0,4) -- (20, 4) node[right] {$s = 4 m_H^2$};
        \begin{scope}
            \clip (-1, -1) rectangle (19.5, 19.5);
            \fill[blue, opacity=0.1]
            ( {-1+4/2*(1+sqrt(1-4/4))}, 4 ) -- 
            plot[domain=4:20] ({-1+\x/2*(1+sqrt(1-4/\x))}, \x) -- 
            ( {-1+20/2*(1+sqrt(1-4/20))}, 20 ) -- 
            ( {-1+20/2*(1-sqrt(1-4/20))}, 20 ) -- 
            plot[domain=20:4] ({-1+\x/2*(1-sqrt(1-4/\x))}, \x) -- 
            cycle;
            \draw[color=blue, domain=4:20] plot ({-1+\x/2*(1+sqrt(1-4/\x))}, \x);
            \draw[color=blue, domain=4:20] plot ({-1+\x/2*(1-sqrt(1-4/\x))}, \x);
            \node[blue, right] (label) at (1.5, 17) {physical region};
        \end{scope}
        \begin{scope}
            \draw[red] (0, 0) node[circle, fill, red, scale=0.3] {} -- (1875/512, 3125/256) node[circle, fill, red, scale=0.3] {}
            -- (7, 12) node[circle, fill, red, scale=0.3] {};
            \node at (3, 3125/256)[above, red]{$(s_0, t_0)$};
            \node at (8.5, 12) [above, red]{$(s, t)$};
        \end{scope}
    \end{tikzpicture}
    \caption{Schematic illustration of the integration path used to evaluate the independent functions.}
    \label{fig:integration_path}
\end{figure}
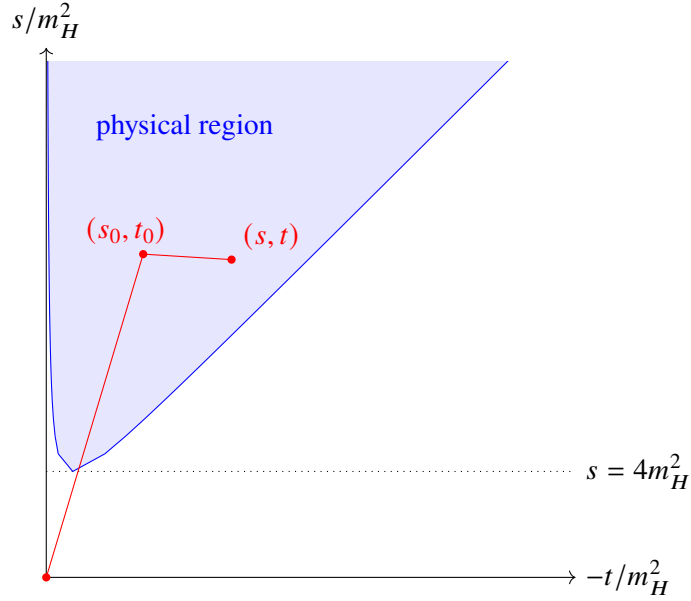

To evaluate the independent functions numerically in the physical region, we solve their differential equations using \texttt{DiffExp} \cite{Hidding:2020ytt}. As an integration path, we choose two straight segments, as schematically shown in Fig.~\ref{fig:integration_path}. The first segment connects the large mass limit to a reference point in the physical region,
\begin{align}
    \left\{s_0, t_0, u_0, m_{H,0}^2, m_{W/Z,0}^2\right\} = \left\{\frac{3125}{256}, - \frac{1875}{512}, - \frac{1875}{512}, \frac{625}{256}, 1\right\}\, .
\end{align}
This part of the path is evaluated once and cross-checked against \texttt{AMFlow} \cite{Liu:2017jxz, Liu:2022chg}. The reference point is then used as a starting point for the second segment to explore the physical region. 

This split of the integration path ensures that all kinematic thresholds, which lie outside the physical region and may otherwise slow down the numerical integration, are crossed only once during the evaluation of the reference point.

The amplitude can be evaluated in this way with a precision of $16$ significant digits in approximately $\mathcal{O}(1')$ per phase space point. As this is not sufficiently fast for a direct implementation into $\texttt{POWHEG-BOX-V2}$ \cite{Nason:2004rx, Frixione:2007vw, Alioli:2010xd}, we construct an interpolation grid comprising $\mathcal{O}(20\, 000)$ phase-space points. Therefore, we compute a rectilinear grid in the parameters 
\begin{align}
    \beta = \sqrt{1 - 4\, \frac{m_H^2}{s}} \in [0,1[ \quad\text{and}\quad\cos\theta = \frac{t - u}{s\, \beta} \in [-1, 1]\,,
\end{align}
with a sampling density weighted according to the amplitude squared. 

The LO amplitudes as well as the real-emission contributions are evaluated on the fly using \texttt{GoSam-3}.

\section{Results}
We calculate differential distributions for a center-of-mass energy of $\sqrt{S} = 13.6\,\mathrm{TeV}$ using the EW input parameters listed in Tab.~\ref{tab:input}. The renormalisation and factorisation scales are chosen dynamically as $m_{HH}/2$, where $m_{HH}$ denotes the invariant mass of the Higgs boson pair. The scale uncertainties are estimated through 7-point scale variations. 

\begin{table}[tb]
    \centering 
    \caption{EW input parameters used for phenomenological results.}
    \label{tab:input}
    \begin{tabular}{cccc}
    \toprule
     $m_H$ [GeV] & $m_W$ [GeV] & $m_Z$ [GeV] & $G_F$ [GeV$^{-2}$] \\ 
    \midrule
     $125$ & $80.36$ & $91.1876$ &  $0.116639\times10^{-4}$\\ 
     \bottomrule
\end{tabular}
\end{table}

\begin{figure}[tb]
    \centering 
    \subfloat[]{\includegraphics[width=0.49\linewidth]{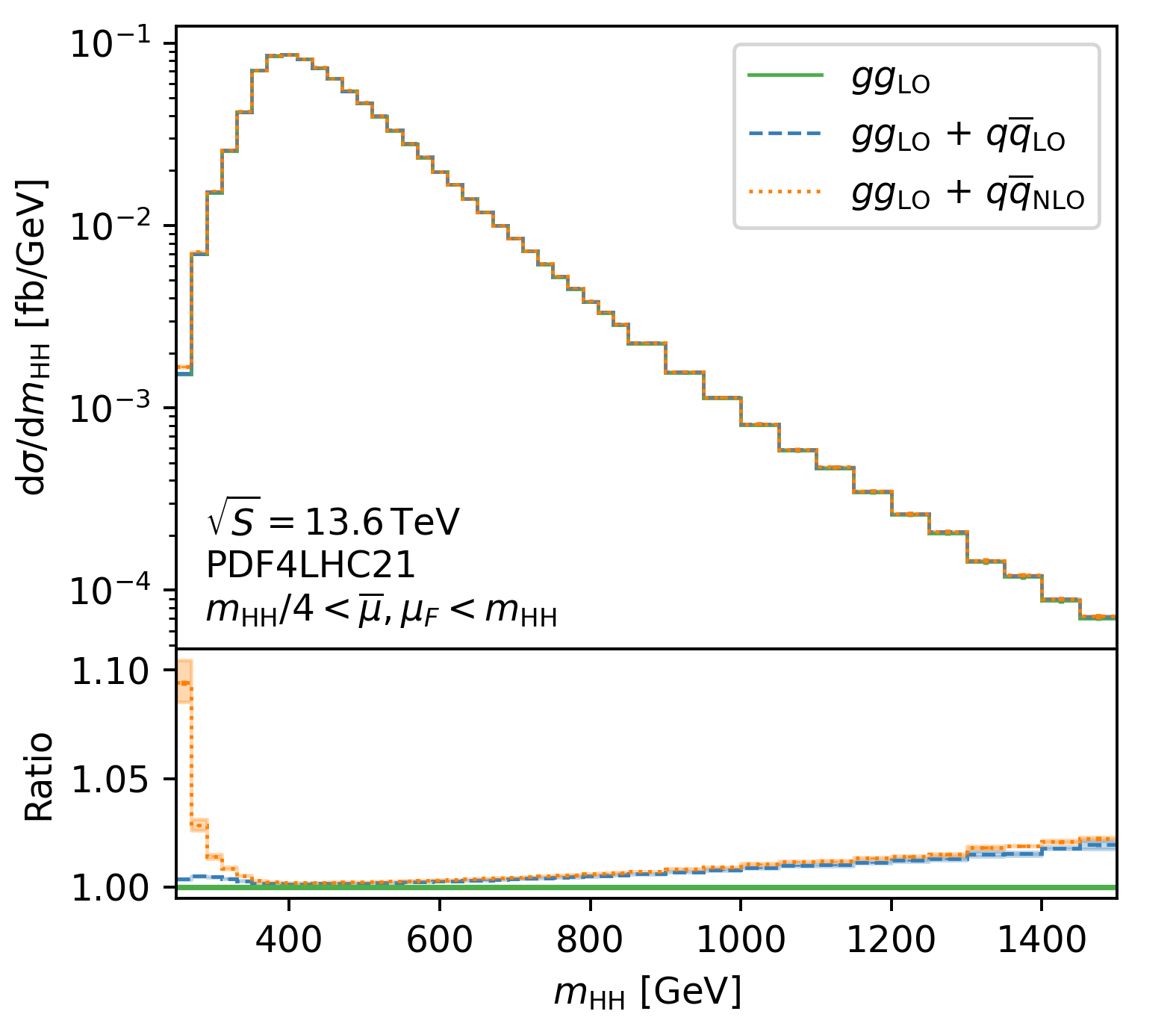}\label{fig:invmass}}
    \subfloat[]{\includegraphics[width=0.49\linewidth]{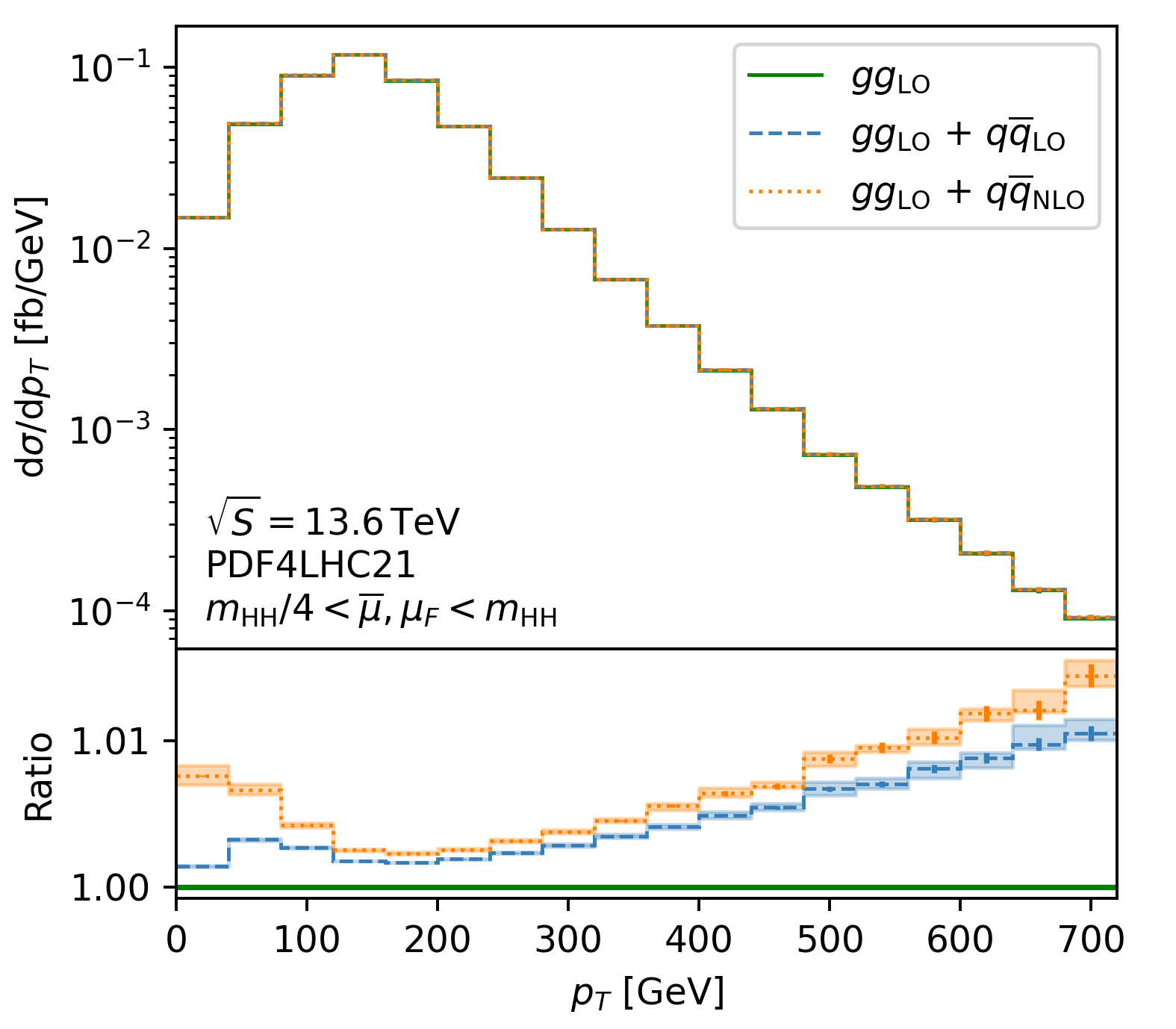}\label{fig:transmom}}
    \caption{Invariant mass distribution of the Higgs boson pair (a) and transverse momentum distribution of a single Higgs boson (b). The error bands represent the scale uncertainty, while the error bars show the statistical error from the Monte Carlo simulation.}
    \label{fig:dist}
\end{figure}

Fig.~\ref{fig:invmass} shows the invariant mass distribution of the Higgs boson pair. The LO and NLO quark--antiquark contributions are shown as dashed blue and dotted orange lines, respectively, together with the LO gluon-fusion channel, represented by the solid green line. The NLO quark--antiquark channel exhibits a significant enhancement close to the production threshold, reaching a relative correction of approximately $+10\%$ compared to LO gluon-fusion. This effect is mainly driven by real-emissions, since the emitted parton softens the spectrum, which leads to a higher population of the threshold region. At larger invariant masses, the NLO corrections remain positive and reach approximately $+3\%$.

Fig.~\ref{fig:transmom} shows the transverse-momentum distribution of one of the Higgs bosons. The corrections exhibit a pattern similar to that observed for the invariant-mass distribution: An enhancement at low transverse momentum, predominantly driven by real-emission contributions, and a moderate increase in the high-$p_T$ tail.

\section{Conclusion}
These proceedings reviewed the calculation of the quark--antiquark channel for Higgs boson pair production up to NLO QCD. The amplitudes were computed analytically, using differential equations to express Feynman integrals in terms of iterated integrals over logarithmic kernels. Boundary conditions were obtained from a large-mass expansion, and additional basis rotations were performed to construct a basis of graded transcendental functions. The resulting amplitudes were evaluated on a grid using \texttt{DiffExp} and implemented into \texttt{POWHEG-BOX} for phenomenological studies. 

We observe that the quark--antiquark channel has a sizeable impact on differential distributions, even though its contribution to the total cross section is negligible. In particular, the invariant-mass distribution of the Higgs boson pair shows an enhancement of nearly $+10\%$ in the threshold region, as well as noticeable Sudakov-type effects at high energies. 

Future work includes the public release of the \texttt{POWHEG} implementation and a detailed study of the interplay between additional EW and QCD contributions to Higgs boson pair production. 

\section*{Acknowledgments}
This work is supported by the \textit{Deutsche Forschungsgemeinschaft} (DFG, German Research Foundation) under grant no.\ 396021762 - TRR 257, and by the Leverhulme Trust, LIP-2021-014.

\bibliographystyle{jhep}
\bibliography{refs}

\end{document}